\documentclass[acmsmall, authorversion, nonacm]{acmart}

\usepackage[capitalize]{cleveref}
\usepackage{xspace}
\usepackage{csquotes}
\usepackage{tikz}
\newcommand{\cupa}{CUPA\xspace}

\let\coopas\cupas

\begin{document}

\title{Towards Computer-Using Personal Agents}

\author{Piero A. Bonatti}
\affiliation{%
  \institution{University of Naples Federico II}
  \city{Naples}
  \country{Italy}
}
\email{pab@unina.it}
\orcid{0000-0003-1436-5660}

\author{John Domingue}
\affiliation{%
  \institution{Knowledge Media Institute, The Open University}
  \city{Milton Keynes}
  \country{UK}
}
\email{john.domingue@open.ac.uk}
\orcid{0000-0001-8439-0293}

\author{Anna Lisa Gentile}
\affiliation{%
  \institution{IBM Research}
  \city{San Jose, CA}
  \country{USA}
}
\email{annalisa.gentile@ibm.com}
\orcid{0000-0002-6401-4175}

\author{Andreas Harth}
\affiliation{%
  \institution{Friedrich-Alexander-Universität Erlangen-Nürnberg \& Fraunhofer Institute for Integrated Circuits IIS}
  \city{Nürnberg}
  \country{Germany}
}
\orcid{0000-0002-0702-510X}
\email{andreas.harth@fau.de}

\author{Olaf Hartig}
\affiliation{%
  \institution{Linköping University}
  \city{Linköping}
  \country{Sweden}
}
\email{olaf.hartig@liu.se}
\orcid{0000-0002-1741-2090}

\author{Aidan Hogan}
\affiliation{%
  \institution{DCC, Universidad de Chile \& IMFD}
  \city{Santiago}
  \country{Chile}
}
\email{ahogan@dcc.uchile.cl}
\orcid{0000-0001-9482-1982}

\author{Katja Hose}
\affiliation{%
   \institution{TU Wien}
   \city{Vienna}
   \country{Austria}
}
\email{katja.hose@tuwien.ac.at}
\orcid{0000-0001-7025-8099}

\author{Ernesto Jimenez-Ruiz}
\affiliation{%
   \institution{City St George's, University of London}
   \city{London}
   \country{UK}
}
\email{ernesto.jimenez-ruiz@city.ac.uk}
\orcid{0000-0002-9083-4599}

\author{Deborah L.\ McGuinness}
\email{dlm@cs.rpi.edu}
\affiliation{%
   \institution{Rensselaer Polytechnic Institute}
   \city{Troy, NY}
   \country{USA}
}
\orcid{0000-0001-7037-4567}

\author{Chang Sun}
\affiliation{%
   \institution{Maastricht University}
   \city{Maastricht}
   \country{The Netherlands}
}
\email{chang.sun@maastrichtuniversity.nl}
\orcid{0000-0001-8325-8848}

\author{Ruben Verborgh}
\affiliation{%
   \institution{IDLab, ELIS, Ghent University -- imec}
   \city{Ghent}
   \country{Belgium}
}
\email{ruben.verborgh@ugent.be}
\orcid{0000-0002-8596-222X}

\author{Jesse Wright}
\affiliation{%
   \institution{Department of Computer Science, University of Oxford}
   \city{Oxford}
   \country{UK}
}
\email{jesse.wright@cs.ox.ac.uk}
\orcid{0000-0002-5771-988X}

\renewcommand{\shortauthors}{}

\begin{abstract}
Computer-Using Agents (CUA) enable users to automate increasingly-complex tasks using graphical interfaces such as browsers. As many potential tasks require personal data, we propose Computer-Using Personal Agents (\coopas) that have access to an external repository of the user's personal data. Compared with CUAs, \coopas offer users better control of their personal data, the potential to automate more tasks involving personal data, better interoperability with external sources of data, and better capabilities to coordinate with other CUPAs in order to solve collaborative tasks involving the personal data of multiple users. 
\end{abstract}

\maketitle

\section{Introduction}

Advances in Generative AI, and particularly Large Language Models (LLMs), have led to the recent release of various \textit{Computer-Using Agents} (\textit{CUAs}) that automatically operate a user's computer on their behalf. These agents use multimodal capabilities to interact with graphical interfaces via simulated mouse and keyboard inputs. Prominent commercial examples of CUAs include OpenAI's Operator, Google's Jarvis, and new functionalities in Anthropic's Claude.

Potential use cases for CUAs involve personal and often sensitive data, such as credit card details for purchases, passport numbers for flight booking, addresses for deliveries, and allergy information for dinner reservations. While modern browsers sometimes store personal data to autocomplete web forms, CUAs could additionally take context into account (e.g., selecting between a home or work address, depending on the purchase) and go beyond simple autocompletion.

Passing personal data to CUAs raises valid concerns about how such data might be (mis)used. Currently, OpenAI's Operator invokes a \textit{takeover mode} for tasks involving sensitive data (e.g., log-in or payment details): the user is required to fill the details in manually~\cite{Operator2025}. Such measures target users' concerns about how their personal information will be used by CUAs. OpenAI themselves state that Operator is \textit{\enquote{still learning, evolving and may make mistakes}}~\cite{Operator2025}. There are thus many open questions relating to the use of personal user data by CUAs.

Conversely, there are many potential benefits to users if CUAs are empowered with personal data. CUAs could autofill forms with personal data for users in a context-aware and potentially generative manner, automating a tedious task. CUAs could potentially enrich personal data with public data to better solve tasks. The CUAs of multiple users could negotiate to achieve a mutually beneficial result based on their users' personal context and preferences.

\begin{figure}
    \centering
    \includegraphics[scale=0.35]{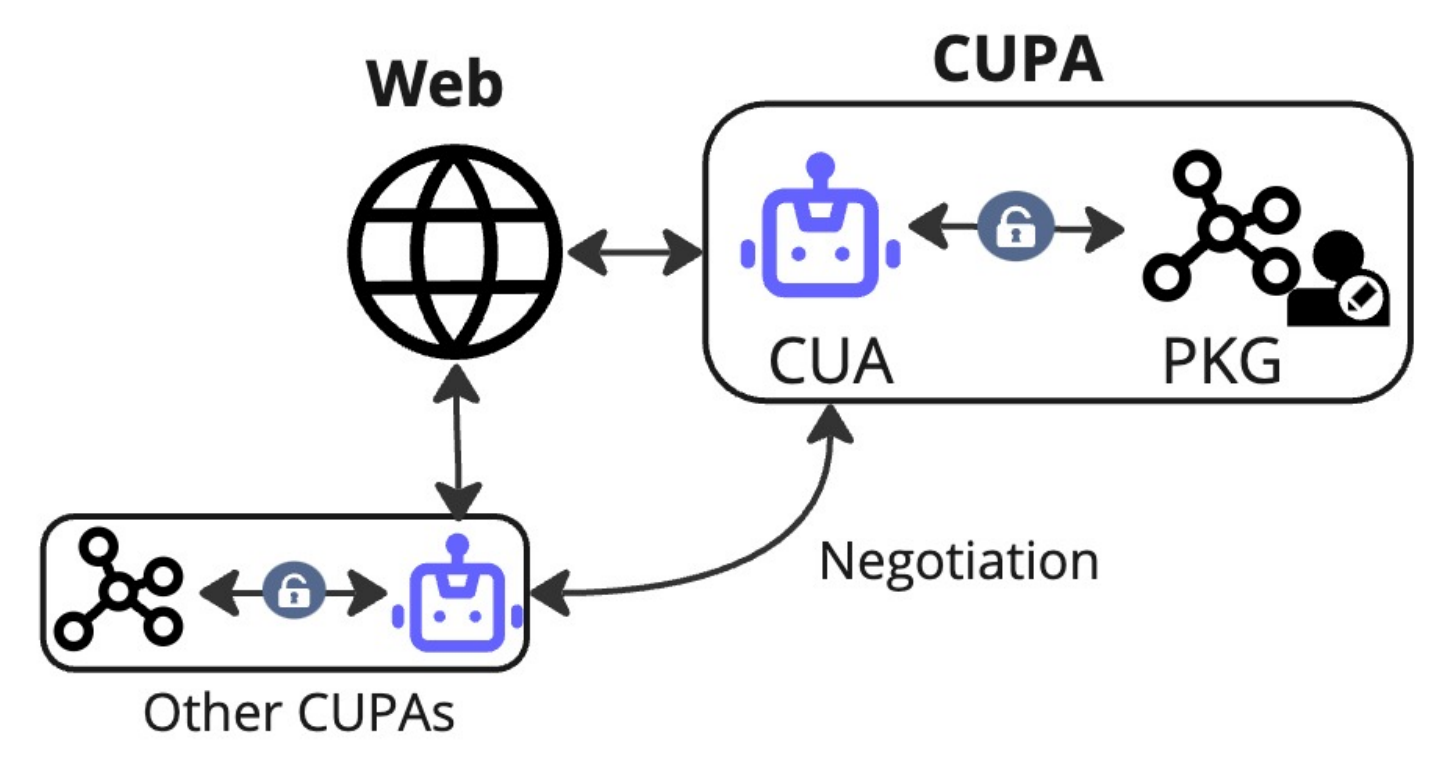}
    \vspace{-\baselineskip}
    \caption{Computer-Using Personal Agent \label{fig:cupa}}
    \Description[Computer-Using Personal Agent]{"The image depicts the flow of Computer-Using Personal Agent."}
\end{figure}

Towards providing users more oversight over their personal data while enabling higher levels of automation for complex tasks, we propose \textbf{Computer-Using Personal Agents} (\textbf{CUPAs}): \textit{a~Computer-Using Agent (CUA) that has controlled access to a structured repository of private information relating to a user}. This concept is illustrated in Figure~\ref{fig:cupa}. Specifically, we propose to instantiate the repository as a \textit{Personal Knowledge Graph} (\textit{PKG}) representing the user's personal data, which would facilitate the specification by users on how the CUA can access and use these data. This PKG can collect more personal data over time, with policies also evolving to reflect the user's fluctuating trust in the system~\cite{afroogh2024trust}.
Looking further forward, one can then imagine a scenario where CUPAs interact with websites and services via the underlying Web APIs instead of through a vision model, where CUPAs can assist in recommendations and negotiations based also on interactions with similar users and/or users' CUPAs. 

We provide a road-map towards realising this vision of CUPAs, discussing what is achievable now with current technology, and what gaps must be addressed via further research and development.

\section{User Scenario}\label{sec:scenarios}

Sam is expecting Jane over for dinner at 8pm, and is thinking about preparing Thai food. Sam is pre-diabetic, while Jane has a shellfish allergy. Sam requests that his CUPA to generates some suggestions of Thai recipes for the occasion. Consulting Sam's schedule, the CUPA recommends to filter recipes requiring more than an hour to prepare based on when he finishes work and his commute time. Sam agrees, and the CUPA starts to retrieve and present shellfish-free recipes of Thai food that are quick to prepare. Upon consulting external sources of nutritional information and recipes  on the Web, the CUPA flags some recipes as being above the postprandial glucose threshold recommended by Sam's doctor (<180 mg/dL), or as having high glycaemic indices (>70).

Sam asks his CUPA to find out what recipe Jane might like. As Sam and Jane are friends, Sam's CUPA can send the candidate recipes to Jane's CUPA to see what she might like. Jane's CUPA suggests to avoid some recipes that include coriander (listed in some recipes as cilantro), which Jane hates. Sam’s agent enforces his glucose thresholds and flags ingredients with high glycaemic indices, using external food and recipe knowledge graphs (e.g., the FoodKG \cite{haussman2019}) to find the alternative ingredients. Of the remaining options, Sam's agent suggests a tofu green curry recipe that catches Sam's eye. Since the recipe is flagged for having a high glycaemic index (78), the agent asks Sam if he might consider replacing jasmine rice with cauliflower rice as a healthier option. Sam refuses the substitution as it is a special occasion.

Sam requests his CUPA to order the ingredients from a local supermarket. Since green bell peppers are out of stock, the CUPA suggests to replace them with yellow bell peppers. Sam agrees, and the CUPA prepares the order for delivery to Sam's home address, soliciting Sam's confirmation. Later that night, Sam and Jane enjoy their dinner of Thai green curry. After Jane leaves, Sam suffers some slight heartburn. He requests his CUPA to order antacids and additionally registers the fact that green curry dishes may cause Sam heartburn for future reference.

\section{State of the Art}\label{sec:soa}

Personal data play an increasingly important role in modern life~\cite{Mortier2016,Birch2021}. Early works~\cite{Jones2012} characterise such data based on the \textit{concept of six senses}: owned by me, about me, directed to me, sent by me, already experienced by me, and useful to me. More restrictive definitions only include data created by the individual~\cite{Bergman2016}, or that the individual cares for/about~\cite{24824,Cushing2022}.

Much literature has been dedicated to Personal Information Management systems (PIMs), which deal with the acquisition, organisation, maintenance, retrieval and sharing of personal data~\cite{jones2007personal}. Notable PIM technologies include blockchain systems~\cite{Zyskind2015,Faber2019}, systems capturing user behaviour on multiple user devices \cite{Lin2022,Schroder2022}, and end-user prototypes~\cite{Chaudhry2015,Mortier2016,Kalokyri2018}. Personal Knowledge Graphs (PKGs)~\cite{Chakraborty2022,Chakraborty2023,skjaeveland2024ecosystem} further apply a graph abstraction to personal data, opening up possibilities for declarative access policies, deductive inference, and integration with external Knowledge Graphs.

Towards taking fuller advantage of such data, AI-powered agents show much promise, particularly those that can automate tasks currently performed by the user. Robotic Process Automation (RPA)~\cite{Aalst,Silva2022} automates interactions with human interfaces. However, such approaches are hard-coded, brittle to changes in the interface, and incapable of generalising to unseen interfaces. Conversely, AI-based agents are capable of learning and generalising. LLM-based agents have been proposed to operate in diverse environments using recursion, feedback, and careful prompt engineering~\cite{YangPNY23}. Such LLM-based agents are capable of solving computer tasks -- despite the limited reasoning capabilities in LLMs~\cite{KimBM23} -- paving the way for CUAs such as Operator~\cite{Operator2025}.

Regarding works unifying LLM-based agents with PKGs, AGENTIGraph ~\cite{Zhao2024} heads in this direction, but rather focuses on question answering. Closer to the idea of CUPAs is Charlie: a brief proposal by Berners-Lee~\cite{charlieWorks} on combining LLM-based agents with PKGs instantiated by Solid pods using Semantic Web standards. This proposal, and the user scenario presented previously, echo the (yet unrealised) vision laid out by Berners-Lee et al.~\cite{berners-lee2001} for the Semantic Web 24 years ago. Wright~\cite{DBLP:journals/corr/abs-2409-04465} presents a ``discuss then transact'' model of LLM-interaction in support of this vision for LLM-based personal agents that represent legal entities.

\section{Added Value}

Societal and legal debates on personal data emphasise \emph{protection} from the harm that they could inflict, and understandably so. Yet people voluntarily exchange personal data with others in their every-day lives in the pursuit of mutual benefit.
People can decide to leverage more personal data, or different kinds of personal data, to achieve a desired outcome. For instance,
patients might prefer to share fitness-tracker data with their doctor if this improves their treatment, or consumers might want to divulge allergies and dietary needs to streamline online shopping and avoid nasty surprises.

A dangerous assumption is that companies are more capable of distilling value from people's personal data than the people the data describe. A company certainly has advantages over individuals in this respect, such as the ability to aggregate over a great many users. But personal data about a particular individual in isolation has much greater potential to empower that individual than a company they interact with, especially when the individual is coached by an agent such as a CUPA. CUPAs representing different parties could even negotiate a better outcome for \emph{all} parties involved.

Considering the added value of CUPAs, and more generally of providing AI-based agents access to personal data, we highlight:

\begin{description}
    \item[Multi-dimensional negotiation.] CUPAs can help users to strike sweet-spots between multiple dimensions, such as the cost and duration of multi-hop flights, the deliciousness and healthiness of meal options, etc.

    \item[Increased granularity.] Humans struggle to negotiate on a fine-grained level, and may thus prefer broad policies that reduce cognitive load (e.g., to always accept all cookies)~\cite{wright2024wantcookieautomatedtransparent}. CUPAs can help to reach fine-grained agreements that improve outcomes and honour party preferences.

    \item[Improved risk/reward assessment.] CUPAs can help users simulate and analyse a variety of hypothetical data exchange scenarios, and warn users of a particular risk, for example that the supermarket -- if informed of a condition of a severe allergy -- could sell this information to third parties, leading to an increase in life assurance premiums.
        
    \item[Auditing and follow-up.] \coopas could automatically perform audits to assess whether the data were treated as agreed during the negotiation process, evaluate the benefit to the user, and improve for future interactions.
\end{description}

\noindent
Such added value is, of course, dependent on the value outweighing the potential harms caused. This can be addressed via AI alignment, which ensures that artificial intelligence systems act in accordance with human intentions, values, and societal norms. It involves \textit{outer alignment}, where an AI’s objectives accurately reflect human goals, and \textit{inner alignment}, ensuring learned behaviours remain aligned in novel scenarios. Machine-readable policies on how personal data from the PKG can or should be used by the AI-based agent can also help to avoid harm. Representing personal data as PKGs allows standards such as the Open Digital Rights Language (ODRL)~\cite{Iannella_Villata_2023} and policy engines implementing formal semantics~\cite{Fornara_Rodríguez-Doncel_Esteves_Steyskal_Smith_2024} to specify and automate the processing of policies about how personal data are used, in what contexts, and under what conditions.

\section{CUPA Capabilities}
\label{sec:capabilities}

\let\capability\emph

Computer-using personal agents must be able to \capability{interact with diverse websites and APIs}. This allows them to book flights and hotels, search for job openings, and even schedule appointments. Moreover, they must possess the ability to \capability{interact with other such agents}, such as coordinating travel arrangements with a travel agent or collaborating with a financial agent to manage expenses.

In addition to being able to \capability{generate and adapt content} (e.g., personalised summaries and creative text), a computer-using personal agent must be able to \capability{combine private data} from the user's personal knowledge graph (PKG) with external information. For example, when searching for a new apartment, the agent should combine the user's preferred neighbourhood from their PKG with data from real estate websites and local amenities databases to find the most suitable options. When utilising the knowledge stored within the PKG, the agent must also be able to \capability{adapt the knowledge from the PKG for the current task}. For instance, when filling out a job application form, the agent should selectively use information from the user's CV and work history stored in the PKG, tailoring the presentation to the specific requirements of each application. This adaptability is crucial for ensuring that agent actions are relevant and effective in the given context. 

CUPAs must continuously \capability{collect and enrich user information} to effectively assist them. This involves gathering data from various sources, including interactions with websites and APIs, user inputs, and external sources. By continuously \capability{learning about user preferences}, these agents can personalise their assistance, such as recommending travel options that align with the user's preferences or suggesting recipes that cater to specific dietary restrictions or tastes. However, it is also crucial for such agents to \capability{avoid learning one-off or irrelevant patterns}, for example, to assume that Sam will always suffer heartburn after eating Thai food and should thus avoid it.

Computer-using personal agents must exhibit a high degree of autonomy. They should ideally \capability{act maximally autonomously}, including the ability to \capability{proactively anticipate and address user needs}. For example, an agent could proactively remind users of upcoming appointments or suggest relevant articles based on their recent reading history. However, this autonomy must always be balanced with the ability to \capability{be guided and controlled by the user}, allowing users to provide instructions, adjust preferences, and maintain control over agentic actions.

While acting largely autonomously, it is crucial that a computer-using personal agent \capability{acts in alignment with the user}, ensuring that tasks are completed as desired. This is essential in scenarios like recipe searches where the agent must accurately reflect dietary restrictions and preferences. Moreover, such an agent should always act in the user's interests, even when \capability{dealing with potentially conflicting goals}. For example, an agent helping a user plan a trip should consider factors like budget, travel time, and personal preferences, even if these factors may lead to a slightly more expensive or less convenient option. The agent should avoid \capability{acting in an unethical or illegal manner} even if it potentially maximises a users immediate interests, e.g., via tax evasion.

To maintain user trust and ensure responsible behaviour, it is also crucial that agents \capability{do not overstep bounds}, respecting user privacy and only acting within explicitly granted permissions. Finally, the repeated offering of \capability{clear explanations of all actions} will aid in the fostering of trust and allow users to understand and verify agent behaviour.

\section{Technical Challenges}

The aforementioned desired capabilities for CUPAs, based on our vision of a trusted, accountable and largely autonomous agent acting with personal data for user benefit, raises a number of technical challenges.

\begin{description}

\item[Accountability and Liability] In the case of undesired, illegal, or unethical acts involving CUPAs, it is important to determine who -- or what -- is responsible, who should be held accountable, and where the liability lies.

\item[Explainability, Traceability, and Provenance] Provenance techniques are required to trace and explain how personal and external data led to specific answers or actions being derived or carried out by the CUPA. These provenance techniques would need to support diverse data models, machine learning processes, user inputs and policies.

\item[Data Interoperability] Data interoperability is a key challenge towards implementing CUPAs. Being able to draw on and integrate more sources of data will improve the CUPAs performance. This is particularly challenging for new sources discovered on the fly.

\item[Inter-Agent Communication, Negotiation and Coordination]%
Agents must communicate effectively in the context of multi-agent systems to achieve shared goals, requiring both a shared conceptual understanding and a means of encoding and decoding messages~\cite{wooldridge2009introduction}. The same challenge applies to networks of CUPAs who coordinate to solve a particular set of goals for users. 

\item[Security, Privacy, and Policies] The sensitive nature of data processed by a \cupa calls for security, privacy, and usage control mechanisms, and the ability of the \cupa to understand and correctly apply the access/usage/action control policies of the user. In some countries, this would even be a legal requirement (e.g., under GDPR in the E.U.).

\item[Trust, Delegation, and Action Control]
Achieving agent autonomy requires trust modelling, delegation mechanisms, and structured action control policies~\cite{south2025authenticated}. Trust models must be adaptable to different contexts, from rigid policies applicable in government agencies to more flexible, reputation-based approaches for personal agents~\cite{chen2015trust}.

\item[User-in-the-Loop]
CUPAs will require input, guidance, permission and confirmation from the user. But to increase automation, the CUPA must avoid unnecessary interactions with the user. This creates the challenge of \textit{when} to call upon the user, and how.

\item[Self-Improvement]
The CUPA should leverage its experience with the user in order to improve the services it provides over time, leading to greater automation, and actions/results that better benefit the user. This raises questions about how such a history can be captured, represented, stored and leveraged.

\item[Self-Determination and Alignment]
There are many ways an agent could be considered `aligned' to a user. Naive approaches include ensuring that CUPA decision making always takes place within rules-based bounds - such as action controls set by a user - or doing a best effort to match user \textit{intent} or \textit{decision making}. There is a field of research discussing `beyond preference matching' alignment which proposes that AI systems should be aligned to broader concepts such as value-based alignment, or prioritise user welfare over emulating decision making~\cite{Zhi_Xuan_2024}.

\end{description}

\section{Roadmap}

We envisage that moving from the current state of the art to fully addressing the above technical challenges will occur in three stages. These levels represent varying degrees of trust, accountability and autonomy.

\begin{description}

\item[CUAs enhanced with personal data]

In the first instance, we foresee extensions of CUAs -- in the style of OpenAI's Operator~\cite{Operator2025} in a commercial setting and Agent-E~\citep{abuelsaad2024agent} in a research setting -- such that they use a PKG in order to access knowledge personal to the user. This would safely enable higher levels of automation, obviating the need to pass control back to the user in scenarios of the user's choosing that involve personal data.

\item[Web-aware CUPAs]

CUAs currently rely on existing browser implementations to render an HTML page and then make use of vision models to interact with the page. An agent could rather observe HTTP requests made to a particular website, as well as the HTML forms present on a page, to invoke requests and actions directly via HTTP.

\item[Networks of CUPAs]

We envision networks of CUPAs interacting in order to complete tasks involving multiple users. This may involve structured service descriptions~\cite{Martin2005}, or a mix of natural language and structured communication per a ``discuss then transact'' model~\cite{DBLP:journals/corr/abs-2409-04465} whereby agents use natural language to first negotiate about a transaction they wish to perform, and then confirm this transaction using structured data.

\end{description}

\section{Conclusion}

Computer-Using Agents (CUAs) have the potential to transform how users interact with their computers, their browsers and amongst themselves. Not having access to personal data limits such interactions. Giving CUAs unfettered access to the personal (and most sensitive) data of a user seems unwise, as does providing CUAs no access to personal data. We thus argue for CUPAs as a configurable middle-ground, where a Personal Knowledge Graph (PKG) is used to represent, store and control access to the personal data of the user. As a starting point, the data that a user fills into web forms can be captured in the PKG, and enriched by an AI-based agent. These data can then be used, if the user so wishes, by CUAs to automate further tasks. In a next step, CUPAs can learn to interact with websites via HTTP APIs rather than though visual interfaces. Finally, we envisage further into the future a network of CUPAs collaborating to address users' tasks.

\begin{acks}
This report is a result of Dagstuhl Seminar 25051 ``Trust and Accountability in Knowledge Graph-Based AI for Self Determination'', which took place in January 2025.
\end{acks}

\bibliographystyle{acm}
\bibliography{pkg}

\end{document}